\documentstyle[12pt,aps,psfig]{revtex}

\begin{document}
\draft
\title{The (11112) model on a 1+1 dimensional lattice}
\author{She-Sheng Xue
}
\address{
ICRA, INFN  and
Physics Department, University of Rome ``La Sapienza", 00185 Rome, Italy
}


\maketitle

\centerline{xue@icra.it}

\begin{abstract}

We study the chiral gauge model (11112) of four left-movers and one right-mover with strong interactions in the 1+1 dimensional lattice. Exact computations of relevant $S$-matrix elements demonstrate a loophole that so constructed model and its dynamics can possibly evade the ``no-go'' theorem of Nielsen and Ninomiya. 

\end{abstract}

\pacs{
11.15Ha,
11.30.Rd, 
11.30.Qc
}

\narrowtext

\section{\it Introduction.}

The parity-violating feature in the low-energy is strongly phenomenologically supported. On the basis of this feature, the successful standard model for particle physics is constructed in the form of a renormalizable quantum field theory with chiral (parity-violating) gauge symmetries. However, with very generic axioms, the ``no-go" theorem \cite{nn81} demonstrates that the quantum field theories with chiral gauge symmetries, as the standard model, cannot be consistently regularized on the lattice for vectorlike fermion doubling phenomenon. Searching for a chiral gauge symmetric approach to properly regularize the standard model on the lattice has been greatly challenging to particle physicists for the last two decades. One of approaches is the modeling 
by appropriately introducing local interactions\cite{ep}-\cite{xue00}. However, it is general belief that the phenomenon of spontaneous symmetry breakings in the intermediate value of couplings and the argument of anomaly-cancelation within vectorlike spectra in the strong coupling region prevent such modelings from having a low-energy scaling region for chiral gauged fermions. Nevertheless, in refs.\cite{xue97}, a chiral gauge model with peculiar interactions and a plausible scaling region were advocated and the dynamics of realizing chiral gauged fermions was intensively studied\cite{xue97l,xue00}. In order to make the model so constructed and its dynamics be more physically intuitive and be easily checked, it is worthwhile to study the (11112) model\footnote{In the overlap formulation, the 
(11112) model was adopted to study a chiral gauge theory on the lattice\cite{onn97}.} with analogous interactions on the 1+1 dimensional lattice. In addition, the features of the parity-violating gauge couplings of neutrinos and other fermions in the standard model can be mimicked by the chiral gauge couplings of the (11112) model. In this article, the exact computations of relevant $S$-matrix elements demonstrate a loophole that the model and its dynamics can possibly evade the ``no-go'' theorem of Nielsen and Ninomiya to achieve low-energy chiral gauged spectra, consistently with both the cancelation of gauge anomalies preserving the gauge symmetry and flavor-singlet anomalies obeying the index theorem. 

The chiral model (11112) is made of a $U(1)$ gauge field, four left-movers $\psi^i_L$ with charge $Q_L^i\equiv Q_L=1$ ($i=1,2,3,4)$ and one right-mover $\psi_R$ with charge $Q_R=2$. The t'Hooft condition for gauge anomaly cancelation $\sum_i(Q^i_L)^2=Q_R^2$ is satisfied.  With the fixed spatial and temporal lattice spacings $a$ and $a_t$ $(a\gg a_t)$, the free Hamiltonian is given by (we henceforth omit the index $i=1,2,3,4$),  
\begin{equation}
H_\circ ={1\over 2a}\sum_x\left(\bar\psi_L(x) D^L\cdot\gamma\psi_L(x)+\bar\psi_R(x) D^R\cdot\gamma\psi_R(x)\right),
\label{free}
\end{equation}
where all fermionic fields are two-component and dimensionless Weyl
fields, $x$ is the integer label of space sites, $\gamma$-matrix ($\gamma^2=1$) and
\begin{eqnarray} 
D^{L,R}&=&([U_1(x)]^{Q_{L,R}}\delta_{x,x+1}
-[U_1^\dagger(x)]^{Q_{L,R}}\delta_{x,x-1}),\nonumber\\
\delta_{x,x\pm 1}\psi_{L,R}(x)&=&\psi_{L,R}(x\pm 1),
\label{kinetic} 
\end{eqnarray} 
where $U_1(x)$ is the gauge field at a spatial link and the temporal gauge fixing $U_\circ(x)=1$. This is a chiral gauge model that cannot be naively quantized on the lattice due to the vectorlike fermion doubling\cite{nn81}. To simplify discussions, we define the physical momentum as $\tilde p$ and dimensionless momentum $p\equiv a\tilde p$.

\section{\it Three-fermion states.}

We introduce two neutral and massless spectators(sterile neutrinos) $\chi_R$ and $\chi_L$. $\chi_R$ couples to four left-movers $\psi^i_L$ and $\chi_L$ couples to the right-mover $\psi_R$ as follow,
\begin{eqnarray}
H^L_i&=&g\sum_x \bar\psi_L(x)\cdot\left[\Delta\chi_R(x)\right]
\left[\Delta\bar\chi_R(x)\right]\cdot\psi_L(x),
\label{hil}\\
H^R_i&=&g\sum_x \bar\psi_R(x)\cdot\left[\Delta\chi_L(x)\right]
\left[\Delta\bar\chi_L(x)\right]\cdot\psi_R(x),
\label{hir}
\end{eqnarray}
where the multifermion coupling $g$ is dimensional
$[a^{-1}]$ and the operator $\Delta$ is given as,
\begin{eqnarray}
\Delta\chi_{L,R}(x)&\equiv&
\left[ \chi_{L,R}(x+1)+\chi_{L,R}(x-1)-2\chi_{L,R}(x)\right],\nonumber\\
w(p)&=&{1\over2}\sum_xe^{-ipx}\Delta(x)=\left(\cos(p)-1\right).
\label{wisf}
\end{eqnarray}
Eq.(\ref{wisf}) indicates that large momentum states of $\chi_R(\chi_L)$ strongly couple to $\psi_L(\psi_R)$, while small momentum states of $\chi_R(\chi_L)$ weakly couple to $\psi_L(\psi_R)$.
For the convenience of computations, we rescale all fermion fields $\psi\rightarrow (a_tg)^{1\over4}\psi$ and rewrite the total Hamiltonian,
\begin{eqnarray}
a_tH={1\over 2a}({a_t\over g})^{1\over2}\sum_x\left(\bar\psi_L(x) D^L\cdot\gamma\psi_L(x)+\bar\psi_R(x) D^R\cdot\gamma\psi_R(x)
+\cdot\cdot\cdot\right)+a_tH^L_i+a_tH^R_i,
\label{action}
\end{eqnarray}
where the coupling $g$ in $H^{L,R}_i$ is rescaled away and ``$\cdot\cdot\cdot$" stands for the kinetic terms for $\chi_{L,R}$. We consider the limit $a_t/a\rightarrow 0$, $ga\rightarrow\infty$ and $ga_t$ is fixed.

This Hamiltonian system (\ref{action}) possesses the continuous the $U(1)$ chiral gauge symmetry, global chiral symmetries $U_{L,R}(1)$ and the shift-symmetries of $\chi_R$ and $\chi_L$\cite{gp}. Due to the Mermin and Wagner theorem\cite{nospon}, these continuous symmetries cannot be spontaneously broken for any values of the coupling $ga$. The shift-symmetries protect the right-mover $\psi_R$ sector and left-movers $\psi_L$ sector from coupling each other\cite{ward} and guarantee the spectators $\chi_R, \chi_L$ decoupled as free particles\cite{xue97,gp}. These features greatly simplify our demonstrations. In this letter, we only take the left-moving sector $\psi_L(x)$ as an example for computing the spectrum of the Hamiltonian (\ref{action}). The computations for the right-mover $\psi_R$ sector are the exactly same as that for the left-moving sector $\psi_L$ and can be obtained by substituting $\psi_L\rightarrow \psi_R$ and $\chi_R\rightarrow \chi_L$. 

For the strong coupling $ga\gg 1$, the three-fermion state $\Psi_R$ with the same quantum numbers of $\psi_L(x)$ is formed\cite{xue97},
\begin{equation}
\Psi_R={1\over
2}(\bar\chi_R\cdot\psi_L)\chi_R,
\label{bound}
\end{equation}
which is a two-component Weyl fermion state. To demonstrate this, we compute the S-matrix element whose pole represents this three-fermion state $\Psi_R(x,t)$,
\begin{equation}
S_{33}(x)\equiv \lim_{t_{f,i}\rightarrow \pm\infty}\langle\Psi_R(0,t_f)|\Psi_R(x,t_i)\rangle,
\label{sbound}
\end{equation}
where
\begin{equation}
|\Psi_R(x,t_i)\rangle=e^{it_iH_i^L}|\Psi_R(x,-\infty)\rangle;\hskip0.3cm
\langle\Psi_R(0,t_f)|=\langle\Psi_R(0,+\infty)|e^{-it_fH_i^L},
\label{asystate}
\end{equation}
and $|\Psi_R(x,\pm\infty)\rangle$ are the asymptotical states. We have then
\begin{equation}
S_{33}(x) = \lim_{t_{f,i}\rightarrow \pm\infty}\langle\Psi_R(0,+\infty)|e^{-it_fH_i^L}e^{it_iH_i^L}|\Psi_R(x,-\infty)\rangle.
\label{ssbound1}
\end{equation}
Using identity ($\psi$ indicates $\psi_L$ or $\chi_R$)
\begin{equation}
\int_ {\psi(x,t)}\bar\psi(x,t)\psi(x,t)\equiv |\psi(x,t)\rangle\langle \psi(x,t)|=1,\hskip0.3cm \int_ {\psi(x,t)}\equiv\int d\psi(x,t)d\bar\psi(x,t),
\label{com}
\end{equation}
to insert $|\psi_L(0,t_f)\rangle\langle \psi_L(0,t_f)|$ and $|\psi_L(x,t_i)\rangle\langle \psi_L(x,t_i)|$ into eq.(\ref{ssbound1}), we obtain
\begin{eqnarray}
S_{33}(x) &=& \lim_{t_{f,i}\rightarrow \pm\infty}\langle\Psi_R(0,+\infty)|e^{-it_fH_i^L}|\psi_L(0,t_f)\rangle\langle \psi_L(0,t_f)|\psi_L(x,t_i)\rangle\nonumber\\
&\cdot&\langle \psi_L(x,t_i)|e^{it_iH_i^L}|\Psi_R(x,-\infty)\rangle.
\label{ssbound}
\end{eqnarray}

We define the form factor of the three-fermion state $\Psi_R(x)$:
\begin{eqnarray}
Z_{t_i}(x)&\equiv&\langle \psi_L(x,t_i)|e^{it_iH_i^L}|\Psi_R(x,-\infty)\rangle
=\Pi_{-\infty}^{t_i}\int_ {\psi(x,t)} e^{-a_tH_i^L}\bar\psi_L(x,t_i)\Psi_R(x,-\infty)\nonumber\\
Z_{t_f}(0)&\equiv&\langle\Psi_R(0,+\infty)|e^{-it_fH_i^L}|\psi_L(0,t_f)\rangle
=\Pi^{+\infty}_{t_f}\int_ {\psi(0,t)} e^{-a_tH_i^L}\bar\Psi_R(0,+\infty)\psi_L(0,t_f).
\label{zz}
\end{eqnarray}
The transfer matrix element $S_{11}(x)$ of two intermediate states $|\psi_L(0,t_f)\rangle$ and $ |\psi_L(x,t_i)\rangle$ is given by
\begin{equation}
S_{11}(x)\equiv\langle \psi_L(0,t_f)|\psi_L(x,t_i)\rangle=\Pi_0^x\Pi_{t_f}^{t_i}\int_ {\psi(y,t)} e^{-a_tH}\bar\psi_L(0,t_i)\psi_L(x,t_f),
\label{tran1}
\end{equation}
where the $H$ is the total Hamiltonian (\ref{action}). In eqs.(\ref{zz},\ref{tran1}), we make the Wick rotation to the Euclidean space. 
As a result, the S-matrix element (\ref{sbound}) can be written as,
\begin{equation}
S_{33}(x)= \lim_{t_{f,i}\rightarrow \pm\infty} Z_{t_i}(x)S_{11}(x)Z_{t_f}(0).
\label{tran}
\end{equation}

From eq.(\ref{zz}), we compute the form factors $Z_{t_i}(0)$ and $Z_{t_f}(x)$: 
\begin{equation}
Z_{t_i}(0)=a_tg\Delta^2(0),\hskip0.5cm Z_{t_f}(x)=a_tg\Delta^2(x).
\label{rzz}
\end{equation}
To compute the transfer matrix element $S_{11}(x)$, we need another transfer matrix element $S_{31}(x)$ of two intermediate states $|\Psi_R(x,t_i)\rangle$ and $ |\psi_L(0,t_f)\rangle $,
\begin{equation}
S_{31}(x)\equiv\langle\psi_L(0,t_f)|\Psi_R(x,t_i)\rangle.
\label{s31}
\end{equation}
The exact recursion relations for $S_{11}(x)$ and $S_{31}(x)$ can be obtained\cite{app} 
\begin{eqnarray} 
S_{11}(x)&=&{1\over a_tg\Delta^2(x)}\left({a_t\over 2a
}\right)^3\sum^\dagger S_{31}(x)\gamma,\label{re11}\\
S_{31}(x)&=&{1\over2}\left({\delta(x)\over 2a_tg\Delta^2(x)}
+{1\over a_tg\Delta^2(x)}\left({a_t\over 2a
}\right)\sum^\dagger S_{11}(x)\gamma\right),
\label{re21}
\end{eqnarray}
where $\sum^\dagger f(x)=f(x+1)-f(x-1)$. These recursion equations can be solved by the Fourier transformations for $p\not=0$,
\begin{eqnarray}
S_{11}(p)&=&\sum_x e^{-ipx}S_{11}(x)
={{ia_t\over 2a}\sin (p)\gamma\over
({a_t\over a})^2\sin^2(p)+M^2(p)},\label{sll21}\\
S_{31}(p)&=&\sum_xe^{-ipx}S_{31}(x) ={{1\over2}M(p)\over
({a_t\over a})^2\sin^2(p)+M^2(p)},\label{slm21}\\
M(p)&=&8agw^2(p),\label{m}
\end{eqnarray}
where $``p''$ is the momentum of the three-fermion state $\Psi_R(p)=\sum_x\Psi_R(x)e^{-ipx}$. For simplifying illustrations in the following, we introduce the notation $\psi(p)$ indicating a fermion $\psi$ at the momentum state ``$p$".

As a result, for $t_{f,i}\rightarrow \pm\infty$ we obtain the exact S-matrix element (\ref{ssbound}) for $|p|\in (0,\pi]$
\begin{eqnarray}
S_{33}(p)
&=&Z(p){{ia_t\over 2a}\sin (p)\gamma\over
({a_t\over a})^2\sin^2(p)+M^2(p)}Z(p),
\label{rsc11}\\
Z(p)&=&4a_tgw^2(p),
\label{rzzp}
\end{eqnarray}
where $Z(p)$ is the form factor (\ref{rzz}) in the momentum space. The pole and its residual(form factor) in the S-matrix element(\ref{rsc11}) at $p\simeq\pi$ represent the three-fermion state $\Psi_R$ (\ref{bound}), which behaves as an elementary particle. We can make a wave-function renormalization of the
three-fermion states $\Psi_R$ at $p\simeq\pi$\cite{xue97l,xue00},
\begin{equation}
\Psi_R|_{ren}=Z^{-1}(\pi)\Psi_R;\hskip1cm Z(\pi)=16a_tg.
\label{rbound}
\end{equation}
$\Psi_R|_{ren}$ mixes with $\psi_L$ to form a 
four-component massive Dirac fermion $\Psi_c=(\Psi_R|_{ren},\psi_L)$, 
represented by the pole of the following S-matrix element,
\begin{eqnarray}
S_c(x)&=&\lim_{t_{f,i}\rightarrow \pm\infty}\langle\Psi_c(0,t_f)|\bar\Psi_c(x,t_i)\rangle,\nonumber\\
S_c(p)&=&{{ia_t\over a}\sin (p)\gamma +M(p)\over
({a_t\over a})^2\sin^2(p)+M^2(p)}={1\over
-{ia_t\over a}\sin (p)\gamma +M(p)},
\label{ds}
\end{eqnarray}
which is obtained by using $S_{11}(x),S_{31}(x)$ and $S_{33}(x)$. This Dirac fermion mass is $8ag/a_t$ obtained at $p\simeq\pi$.

The bilinear Hamiltonian $H_{\rm bound}$ on the basis of the eigen-states of $\psi_L$ and three-fermion state $\Psi_R$, i.e., Dirac fermion $\Psi_c$, can be obtained from eq.(\ref{ds}) for $p\not=0$:
\begin{equation}
-a_tH_{\rm bound}=-{ia_t\over a}\sin (p)\gamma +M(p),\hskip0.3cm H_{\rm bound}={i\over a}\sin (p)\gamma -{M(p)\over a_t}.
\label{eff}
\end{equation}
This clearly indicates that the negative binding energy of the three-fermion state $\Psi_R$ is ${-M(p)/a_t}$.
Due to the locality of the
model proposed, the vectorlike spectrum (\ref{eff})
obtained by the strong coupling for large momentum states $p\simeq\pi$, can be analytically
continued to small momentum states $p\simeq 0$. However, eq.(\ref{ds}) does not have a massless pole at $p\simeq 0$, since the form factor $Z(p)$ positively vanishes as $O(p^4)$ and the S-matrix element $S_{33}(p)$ (\ref{rsc11}) is no longer singular at $p=0$\cite{note}. Beside, we are not allowed to make wave-function renormalization (\ref{rbound}) for the inverse form factor $Z^{-1}(0)$ being divergent. What happens then to the three-fermion state $\Psi_R$ when $p\rightarrow 0$? 

\section{\it Three-fermion cut.}

As a bound state,
$\Psi_R(p)$ consists of three fermion constituents $\psi_L(-q')$, $\chi_R(p')$ and $\bar\chi_R(q)$ with the total momentum $p=p'+q-q'$, where $|q|\sim\pi$, $|q'|\sim\pi$ and $|p'|\sim\pi$. Two of constituents $\bar\chi_R(q)$ and $\psi_L(-q')$ form a ``soft'' pair ($|q-q'|\ll\pi $) that couples to the third constituent $\chi_R(p')$ so that their relative momentum $|\Delta p|=|p'-(q-q')|\sim |p'|\sim |p|$. The Heisenberg uncertainty principle gives $|\Delta p||\Delta x|\sim 2\pi a$, where $|\Delta x|$ is the size of the bound state $\Psi_R(x)$. If $|\Delta p|\sim |p| \sim\pi$, the size $|\Delta x|$ is the order of the lattice spacing $a$, indicating that $\Psi_R(x)$ is well localized. However, as  
``$|p|$" goes to zero, the size $|\Delta x|$ increases, indicating that $\Psi_R(x)$ spreads over space and dissolves into its three fermion constituents -- a {\it three-fermion cut}\cite{xue97l,xue00}.

This dissolving is demonstrated by the analytical continuation of the momentum $|p|$ in the S-matrix element (\ref{rsc11}) to zero\cite{smatrix}. The form factor $Z(p)$ (\ref{rzzp}), which is inverse proportional to the size $|\Delta x|$, positively vanishes for $|p|\rightarrow 0$ in such a way that the S-matrix element (\ref{rsc11}) does not have a pole at $\tilde p=0$. The necessary condition of the dissolving phenomenon to occur is that for $|p|\sim 0$, the negative binding energy ${-M(p)/a_t}$ vanishes\cite{ssb} and reaches an energy-threshold of the three-fermion cut.
     
The three-fermion cut ${\cal C}[\Psi_R]$ is a virtual state rather than a particle state. For a given total momentum ``$\tilde p$" in the low-energy region, this virtual state ${\cal C}[\Psi_R](\tilde p)$ has not only the same total momentum ``$\tilde p$" as $\Psi_R(\tilde p)$, but also contains the same constituents as $\Psi_R(\tilde p)$: 
$\psi_L(-\tilde q')$, 
$\chi_R(\tilde p')$ and $\bar\chi_R(\tilde q)$, where $\tilde p=\tilde p'+\tilde q-\tilde q'$. These constitutes are low-energy excitations for $a\tilde p'\sim 0, a\tilde q\sim 0$ and $a\tilde q'\sim 0$. In addition,
the three-fermion cut ${\cal C}[\Psi_R]$ has the same quantum numbers as the bound state $\Psi_R(x)$ so that this dissolving phenomenon is chiral gauge symmetric.

The total energy $E_t$ of such a virtual state ${\cal C}[\Psi_R]$ is given by
\begin{eqnarray}
E_t&=&E_1(\tilde p)+E_2(\tilde q)+E_3(\tilde q'),\nonumber\\
E_1(\tilde p')&=& \tilde p'>0,\hskip0.2cm \tilde p'>0 \hskip0.2cm {\rm for} \hskip0.2cm \chi_R, \nonumber\\ 
E_2(\tilde q)&=& -\tilde q>0,\hskip0.2cm \tilde q<0 \hskip0.2cm {\rm for} \hskip0.2cm \bar\chi_R, \nonumber\\ 
E_3(\tilde q')&=&-\tilde q'>0,\hskip0.2cm \tilde q'<0 \hskip0.2cm {\rm for} \hskip0.2cm \psi_L. 
\label{totale}
\end{eqnarray}
Unlike the dispersion relation (\ref{eff}) for the three-fermion state $\Psi_R$, there is no any definite one to one relationship between the total energy $E_t$ and the total momentum ``$\tilde p$" of the virtual state ${\cal C}[\Psi_R](\tilde p)$. Instead, the total energy spectrum $E_t$ of the virtual state ${\cal C}[\Psi_R]$ is continuum with respect to the given total momentum ``$\tilde p$".  
By minimizing $E_t$ (\ref{totale}) with the condition $\tilde p=\tilde p'+\tilde q-\tilde q'$, we obtain the lowest energy $E^{\rm min}_t(\tilde p)$ of the virtual state ${\cal C}[\Psi_R]$,
\begin{equation}
E^{\rm min}_t(\tilde p)=|\tilde p|\le E_t= |\tilde p'|+|\tilde q|+|\tilde q'|.
\label{thre} 
\end{equation}

Given the same total momentum ``$\tilde p$" in the low-energy region, the three-fermion state $\Psi_R(\tilde p)$ is stable, only if only there is an energy gap $\Delta(\tilde p)$ between the lowest energy (\ref{thre}) of the three-fermion cut ${\cal C}[\Psi_R](\tilde p)$ and the negative binding-energy (\ref{eff})
of the three-fermion state $\Psi_R(\tilde p)$, i.e., 
\begin{equation}
\Delta(\tilde p)= E^{\rm min}_t(\tilde p)-(-{M(\tilde p)\over a_t})>0.
\label{sta}
\end{equation}
As the energy gap $\Delta(\tilde p)$ vanishes for $a\tilde p\rightarrow 0$, the three-fermion state $\Psi_R(\tilde p)$ must dissolve into its virtual state ${\cal C}[\Psi_R](\tilde p)$. $\Delta(\tilde p)=0$ determining the critical momentum-threshold $|\tilde p_c|$ for the dissolving phenomenon in the low-energy limit,
\begin{equation}
|\tilde p_c|={1\over a}\left({a_t\over 2a^2g}\right)^{1\over3}\ll {\pi\over2a},\hskip0.3cm ag\gg 1, \hskip0.3cm{a_t\over a}\ll 1,
\label{con}
\end{equation}
and the critical energy-threshold $\epsilon_c=\sqrt{2}|\tilde p_c|$. As an example, for $ag=100$ and $a_t/a=10^{-1}$, we have $|\tilde p_c|\simeq 0.1a^{-1}$ and $|\epsilon_c|\simeq
0.1a^{-1}$. $\pm\epsilon_c$ will be considered just below and above the zero-energy level $E=0$. The dissolving region $|\tilde p|<\tilde p_c$ and $|\epsilon|<\epsilon_c$ is schematically sketched as a ``box'' in 
the energy-momentum $E-p$ plane (see Fig.(\ref{epf}). Given the ``total'' momentum $\tilde p\in (-\tilde p_c,\tilde p_c)$,
the ``total'' energy $E_t$ is the continuous spectrum in between $(-\epsilon_c,\epsilon_c)$, indicating (see eqs.(\ref{totale})) that $E_1(\tilde p')= \tilde p'$, $E_2(\tilde q)=-\tilde q$ and $E_3(\tilde q')=-\tilde q'$ are spectra in between $(-\epsilon_c,\epsilon_c)$ with $\tilde p=\tilde p'+\tilde q-\tilde q'$. This means that $\chi_R(\tilde p')$, $\bar\chi_R(\tilde q)$ and $\psi_L(\tilde q')$ are low-energy modes in $[0,\epsilon_c)$, and  fully fill the negative energy-levels in the $(0,-\epsilon_c]$.

\section{\it Fermion spectrum flow and anomalies.}

\begin{figure}[t]
\centerline{
\psfig{figure=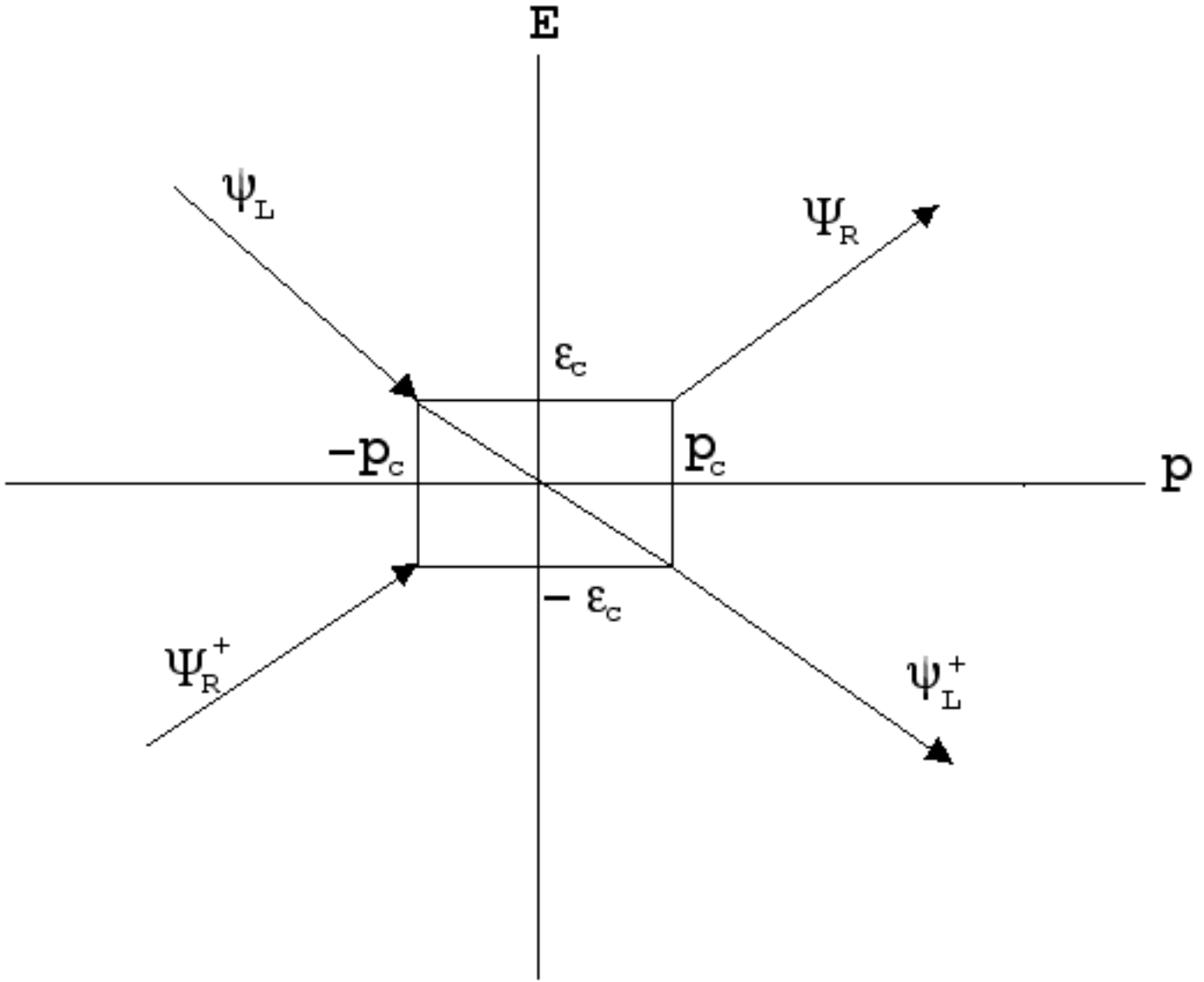}
}
\caption[]{The schematic sketch of the energy-momentum ($E-p$) plane.}
\label{epf}
\end{figure}

The lattice vacuum of the free Hamiltonian is completely modified by the strong interactions (\ref{hil},\ref{hir}). 
We discuss how the lattice vacuum, represented by the energy-momentum $E-p$ plane (see Fig.(\ref{epf}), is filled by the fermion spectrum. In the region $\pi\ge |p|>a|\tilde p_c|$ (outside of the ``box'' in Fig.(\ref{epf}), $\psi_L(-p)$ mixes up with $\Psi_R(p)$ to form the Dirac fermion $\Psi_c(p)$ (\ref{ds}), indicated by the dispersion relation:
\begin{equation}
E=\pm {1\over a_t}\sqrt{({a_t\over a})^2\sin^2(p)+M^2(p)},
\label{dis}
\end{equation}
which is schematically sketched in Fig.(\ref{epf}). This means that its Weyl-components $\psi_L(-p)$ and $\Psi_R(p)$ fill into the same positive energy state $(E>0)$, while $\bar\psi_L(p)$ and $\bar\Psi_R(-p)$ fill into the same negative energy state $(E<0)$. Thus, each energy state in $\pi\ge |p|>a|\tilde p_c|$ is filled by both left- and right- movers with the same quantum numbers so that chiral symmetries is preserved by the vectorlike spectrum $(\psi_L(-p),\Psi_R(p))$.

In the region $|p|\le a|\tilde p_c|$, i.e., $-\epsilon_c<\tilde p<\epsilon_c$ (inside of the ``box'' in Fig.(\ref{epf}) where three-fermion states both the $\Psi_R(p)$ of the $\psi_L$-sector and $\Psi_L(p)$ of the $\psi_R$-sector dissolve into three-fermion cuts, the low-energy spectra are charged fermions $\psi_L^i,\psi_R$ and free neutral fermions $\chi_R,\chi_L$. The neutral fermions $(\chi_L,\chi_R)$ form a free and massless Dirac particle. The energy states $E\in [0,-\epsilon_c)$ are only filled by the charged left(right)-movers $\psi^i_L(-\tilde p)(\psi_R(\tilde p))$ without their partners $\Psi_R(\tilde p)$($\Psi_L(-\tilde p)$) of the same charge and opposite chiralities. Thus the $U(1)$ chiral gauge symmetry and global chiral symmetries $U_{L,R}(1)$ are broken and anomalies must appear, which will be discussed soon.   

The gauge potential $A_\circ=0$ and $A_1$ give the ``electric'' field ${\cal E}=\partial_\circ A_1$ in the direction of $p>0$ to the right-handed side. The ``electric'' force $\pm Q{\cal E}$, sign ``$+(-)$" for the right(left)-mover along(against) the $p$-direction, is the rate of changing momentum states $``p"$ (along the dispersion relation in the phase space) by the unit of $2\pi$ per unit volume of space-time\cite{nn91}, which can be easily understood by the equation of motion and the Heisenberg uncertainty principle $\Delta x\Delta\tilde p\simeq 2\pi$ for the unit quantum $2\pi$ of momentum states. Therefore, $\pm Q^2{\cal E}/4\pi$ is the rate of flowing quantum charges (along the dispersion relation in the phase space) per unit volume of space-time. Note that the factor $1/2$ in the normalization per the unit quantum is owing to two-components of a Weyl field.

In the region $\pi\ge |p|>a|\tilde p_c|$, the ``electric'' force drives both four left-movers $\psi^i_L(p)$ and corresponding three-fermion fermions states $\Psi_R^i(p)$ flowing along their dispersion relations (\ref{dis}) in the energy-momentum $E-p$ plane. In the region $-\pi\le p\le -a|\tilde p_c|$ and for an example the $\psi_L^i$-sector, the ``electric'' field ${\cal E}$ pushes four left-movers $\psi^i_L(p)$ down to the energy-threshold $\epsilon_c$ and pumps four three-fermion states $\bar\Psi^i_R(p)$ up to the energy-threshold $-\epsilon_c$. In the region $\pi\ge p\ge a|\tilde p_c|$, the ``electric'' field ${\cal E}$ pumps four three-fermion states $\Psi^i_R(p)$ away from the energy-threshold $\epsilon_c$ and pushes four left-movers $\bar\psi^i_L(p)$ down below the energy-threshold $-\epsilon_c$.
The rate of pumping out four three-fermion states $\bar\Psi^i_R(p)$ in the region $-\pi\le p<-a|\tilde p_c|$ is the exactly same as the rate of pushing down four left-movers $\bar\psi^i_L(p)$ in the region $\pi\ge p>a|\tilde p_c| $, consistently with (i) the total number of states of the lattice vacuum being finite and (ii) the net-charge flowing being zero in each energy-level, which is in fact the charge-conservation (gauge invariance) of the lattice vacuum. The discussions are the same for $(\psi_R,\Psi_L)$.    
 
In the region $|p|\le a|\tilde p_c|$, as discussed, the energy states $E\in [0,-\epsilon_c)$ are only filled by the charged left(right)-movers $\psi^i_L(-\tilde p)(\psi_R(\tilde p))$, since their partners $\Psi_R(\tilde p)$($\Psi_L(-\tilde p)$) of the same charge and opposite chiralities dissolve. Therefore, the flowing charges near to the zero-energy level ($E=0$) are only contributed from the ``electric'' field pushing down four left-movers $\bar\psi^i_L(p)$ into the lattice vacuum. The corresponding rate of pushing down the charge $Q_L$ of four left-movers $\bar\psi^i_L(p)$, crossing the zero-energy level per unit quantum and per unit volume of space-time, into the lattice vacuum is,   
\begin{equation}
-(Q_L)^2{\cal E}/(4\pi).
\label{gal}
\end{equation}
This is nothing, but by the definition, the divergence of the gauge current $J^i_L=Q_L\bar\psi^i_L\gamma\psi^i_L$, i.e., the gauge anomaly of the $\psi_L^i$-sector. Eq.(\ref{gal}) does not vanish, implying charge non-conservation (gauge variance) of the lattice vacuum. 
Analogous discussions for the $\psi_R$-sector lead to the result that the gauge anomaly associating to the divergence of the gauge current $J_R=Q_R\bar\psi_R\gamma\psi_R$ is,
\begin{equation}
+ (Q_R)^2{\cal E}/(4\pi).
\label{gar}
\end{equation}
Analogously, this is the rate of pumping up the charge $Q_R$ of one right-mover $\psi_R$, crossing the zero-energy level $E=0$, out from the lattice vacuum. Since the t'Hooft condition is obeyed, the rate (\ref{gal}) of pushing down the $U(1)$-charges $Q_L$ into the lattice vacuum is the exactly same as the rate (\ref{gar}) of pumping up the $U(1)$-charge $Q_R$ out from the lattice vacuum. The $U(1)$ {\it net-charge} crossing the zero-energy level $E=0$ per unit volume of space-time, pushed down into and pumped out from the lattice vacuum by the ``electric'' field ${\cal E}$, is identically zero. In the other words, the gauge anomalies are exactly canceled between the $\psi^i_L$-sector and the $\psi_R$-sector. The $U(1)$ charges are conserved, i.e., the $U(1)$ gauge symmetry is preserved by the lattice vacuum. Thus, the gauge invariant model and dynamics discussed in this letter are self-consistent.

However, the corresponding {\it net-number} of zero modes crossing the zero-energy level $E=0$ per unit volume of space-time, pushed down into and pumped out from the lattice vacuum by the ``electric field'', is not zero, i.e., 4-1=3. It seems to be inconsistent for the finiteness of total number of fully occupied states of the lattice vacuum, since there are no extra rooms for accommodating 3 zero modes! 

Based on the discussions leading to the physical meaning of eqs.(\ref{gal},\ref{gar}), we directly obtain that the divergences of flavor-singlet currents $j_L=\sum_i\bar\psi^i_L\gamma\psi^i_L$ and for $j_R=\bar\psi_R\gamma\psi_R$, i.e., the flavor-singlet anomalies, are respectively given by 
\begin{equation}
-{Q_L{\cal E}\over(4\pi)}\hskip0.2cm {\rm and }\hskip0.2cm +{Q_R{\cal E}\over(4\pi)},
\label{sga}
\end{equation}
which are related to ``electric forces'' acting on the left-moves $\psi^i_L$ and right-mover $\psi_R$. As discussed, $-{Q_L{\cal E}\over(4\pi)}$($+{Q_R{\cal E}\over(4\pi)}$) is the rate of pushing down(pumping up) the number $n_-(n_+)$ of state of left(right)-movers down into(out from) the lattice vacuum by crossing the zero-energy level ($E=0$). The corresponding {\it net-number} of zero modes crossing the zero-energy level ($E=0$) $\Delta n=n_--n_+=3$ per unit quantum and unit volume of space-time. On the other hand,  the index theorem requires that the axial anomaly of the axial current $j^5=j_R-j_L$ is given by $\Delta n_{\rm index}=n_+-n_-=-3$. $\Delta n_{\rm index}$ is the numbers of fermionic zero-modes carried by topological gauge fields. $\Delta n_{\rm index}=-3$ indicating three zero modes flowing out from the lattice vacuum, and as a result, three states in the lattice vacuum are emptied for accommodating 3 zero modes pushed into, consistently with the finiteness of the total number of states of the lattice vacuum. Putting it in a different way, we can say that the finiteness of the total number of states of the lattice vacuum results in the index theorem. 

In fact, for the reason that the dissolving energy-scale $\epsilon_c\ll\pi/a\ll\pi/a_t$, the asymmetry $a\gg a_t$ in the space-time turns out to be irrelevant at the low-energy scale $\epsilon_c$.
We can define a low-energy effective Lagrangian for the (11112) model with a continuous regularization at the energy scale $\epsilon_c$.
The asymmetry of filling the left-movers $\psi_L^i$ and right-mover $\psi_R$ into the energy states $E\in [0,-\epsilon_c)$ without their partners of the same charge but opposite chiralities, must appear as explicit chiral-symmetry breaking terms in the low-energy effective Lagrangian at the scale $\epsilon_c$. As results, we have, (i) the gauge anomalies $ -(Q_L)^2\epsilon^{\mu\nu}\partial_\mu A_\nu/(4\pi)$ and $+(Q_R)^2\epsilon^{\mu\nu}\partial_\mu A_\nu/(4\pi)$ of the $U(1)$ gauge symmetry; (ii) the flavor-singlet anomalies $-Q_L\epsilon^{\mu\nu}F_{\mu\nu}/(4\pi)$ and $+Q_R\epsilon^{\mu\nu}F_{\mu\nu}/(4\pi)$ of the $U_{L,R}(1)$ global chiral symmetries and the axial anomaly given by the index theorem; (iii) local counterterms of explicit chiral symmetry breakings at the scale $\epsilon_c$, which can be eliminated by the normal prescription of renormalizable and perturbative quantum field theories.

\section{\it Conclusion.}
 
(i) A chiral gauge symmetric (11112) model with strong interactions is proposed on the 1+1 dimensional lattice, and we exactly demonstrate the dynamics realizing the chiral spectrum in the low-energy region and vectorlike spectrum in the high-energy region. (ii) We show that the cancelation of gauge anomalies is consistent with the vanishing net-charge crossing the zero-energy level $E=0$ of the lattice vacuum, while the non-vanishing net-number of zero-modes crossing the zero-energy level ($E=0$) of the lattice vacuum is consistently canceled by zero-modes carried by gauge fields as required by the index theorem. As a result, the lattice vacuum is chiral symmetric and fully filled and there are no extra zero modes. We conclude that this model with strong interactions on the 1+1 dimensional lattice shows a loophole that can possibly evade the ``no-go'' theorem of Nielsen-Ninomiya. It also provides a simple model for both analytically and numerically verifying the dynamics occuring in the low-energy scaling region\cite{xue97,xue00} for chiral gauge theories. In particular, the numerical verification is very inviting. In addition, the feature of the chiral fermion content of the (11112) model is analogous to that of the standard model for leptons and quarks. We have studied the analogous formulation of the standard model on the lattice\cite{xuesm} and preliminary results seem promising, provided that hard spontaneous symmetry breakings $1/a>\epsilon_c$ is not tolerated and the soft spontaneous symmetry breaking $m\ll\epsilon_c$ is allowed for the fermion mass generation. All these studies strongly imply that the parity symmetry is restored by vectorlike spectra and gauge couplings in the high-energy region.


\newpage 


\begin{references}
\bibitem{ward}
This property can be exactly demonstrated by the Ward identity associating with the shift symmetry.
\bibitem{app} 
See the appendix C of ref.\cite{xue97}, eqs.(C15), (C17) and (C18) are the useful identities and equations for integrating Grassmann variables.
\bibitem{note}
It should be mentioned that if the operator $\Delta=1$ in the interacting Hamiltonian $H_i^L$ in eq.(\ref{hil}), the same computations lead to the bound states, e.g., $\Psi_R(x)$ with the binding energy $-2ag/a_t$ and the form factor $Z=a_tg$ at $p=0$, indicating a massive Dirac fermion at $p=0$.
\bibitem{ssb}
Had hard spontaneous symmetry breakings $m\sim O(1/a)$ developed in an intermediate value of the effective coupling $ag(p)$, the energy-mass gap would have never vanished.
\end{references}
\end{document}